\documentclass[journal]{IEEEtran}
\usepackage{ifpdf}

\usepackage{cite}
\usepackage{hyperref}

\ifCLASSINFOpdf
  \usepackage[pdftex]{graphicx}
\else
\fi
\usepackage{amsmath,amssymb}
\usepackage{array}
\usepackage{threeparttable}
\begin{document}
%
\title{Technology Aware Training in Memristive Neuromorphic Systems based on non-ideal Synaptic Crossbars }
%
%
%

\author{Indranil~Chakraborty, Deboleena~Roy
        and~Kaushik~Roy,~\IEEEmembership{Fellow,~IEEE}
\thanks{I~Chakraborty, D~Roy and K~Roy are with the Department
of Electrical and Computer Engineering, Purdue University, West Lafayette,
IN, 47906 USA e-mail: ichakra@purdue.edu;roy77@purdue.edu;kaushik@purdue.edu}}
\maketitle

\begin{abstract}
The advances in the field of machine learning using neuromorphic systems have paved the pathway for extensive research on possibilities of hardware implementations of neural networks. Various memristive technologies such as oxide-based devices, spintronics and phase change materials have been explored to implement the core functional units of neuromorphic systems, namely the synaptic network, and the neuronal functionality, in a fast and energy efficient manner. However, various non-idealities in the crossbar implementations of the synaptic arrays can significantly degrade performance of neural networks and hence, impose restrictions on feasible crossbar sizes. In this work, we build mathematical models of various non-idealities that occur in crossbar implementations such as source resistance, neuron resistance and chip-to-chip device variations and analyze their impact on the classification accuracy of a fully connected network (FCN) and convolutional neural network (CNN) trained with standard training algorithm. We show that a network trained under ideal conditions can suffer accuracy degradation as large as 59.84\% for FCNs and 62.4\% for CNNs when implemented on non-ideal crossbars for relevant non-ideality ranges. This severely constrains the sizes for crossbars. As a solution, we propose a technology aware training algorithm which incorporates the mathematical models of the non-idealities in the standard training algorithm. We demonstrate that our proposed methodology achieves significant recovery of testing accuracy within 1.9\% of the ideal accuracy for FCNs and 1.5\% for CNNs. We further show that our proposed training algorithm can potentially allow the use of significantly larger crossbar arrays  of sizes 784$\times$500 for FCNs and 4096$\times$512 for CNNs with a minor or no trade-off in accuracy. \end{abstract}


%
\IEEEpeerreviewmaketitle

\section{Introduction}
%
%
%
%
\IEEEPARstart{R}{ecent}
developments in computational neuroscience have resulted in a paradigm shift away from Boolean computing in sequential von-Neumann architectures as the research community strives to emulate the functionality of the human brain on neurocomputers. Although extensive research has been done to accelerate computational functions such as matrix operations on general-purpose computers, the parallelism of the human brain has remained elusive to von-Neumann architecture, thus engendering high hardware cost and energy consumption\cite{Wulf_1995}. This has resulted in the exploration of non-von Neumann architectures with `massively parallel operations in-memory', thus avoiding the overhead cost of exchanging data between memory and processor. Especially with the recent advances in machine learning in various cognitive tasks such as image recognition, natural language processing etc, the search for such energy-efficient `in-memory computing' platforms has become quintessential. Although standardized hardware implementations of neuromorphic systems like $CAVIAR$\cite{Serrano_Gotarredona_2009}, $IBM$ $TrueNorth$\cite{Merolla_2011}, $SpiNNaker$\cite{Xin_Jin_2010} have primarily been dominated by CMOS technology, the memristor-based non-volatile memory (NVM) technology\cite{chua1971memristor,shiga20101,osada2005phase,sheu20114mb,chung2010fully} has naturally evolved into an exciting prospect. To that end, various technologies such as spintronics\cite{Sengupta_2016}, oxide-based memristors \cite{Prezioso_2015,Liu_2016}, phase change materials (PCM) \cite{Eryilmaz_2014}, etc., have shown promising progress in mimicking the functionality of the core computational units of a neural network, i.e., neurons and synapses. \par
The core functionality of a neuromorphic system is a parallelized dot product between the inputs and the synaptic weights\cite{Schmidhuber_2015}. This has been demonstrated to be efficiently realized by a dense resistive crossbar array \cite{Peng_Gu_2015,Sengupta2_2016}. The ability to naturally compute matrix multiplications makes crossbar arrays the most convenient way of implementing neuromorphic systems. However, real crossbars could suffer from various non-idealities including device variations\cite{Yu_2011,Kim_2012}, parasitic resistances, non-ideal sources, and neuron resistances. Although neural networks are generally robust against small variations in the crossbar, the aforementioned technological constraints can severely impact accuracy of recognition tasks as well as restrict the crossbar size. Several techniques such as redundancy schemes\cite{Kannan_2014}, technology optimization\cite{Chen2_2015} and modified training algorithms\cite{Liu_2014,Chen_2015,Liu_2017} have been explored for both on-chip and ex-situ learning to mitigate specific non-ideal effects such as IR drops, synaptic device variations. However, mathematical modeling of non-idealities and its incorporation in standard training algorithm needs further exploration. \par 
\begin{figure*}[t]
		\centering
		\includegraphics[width=6.6in,keepaspectratio]{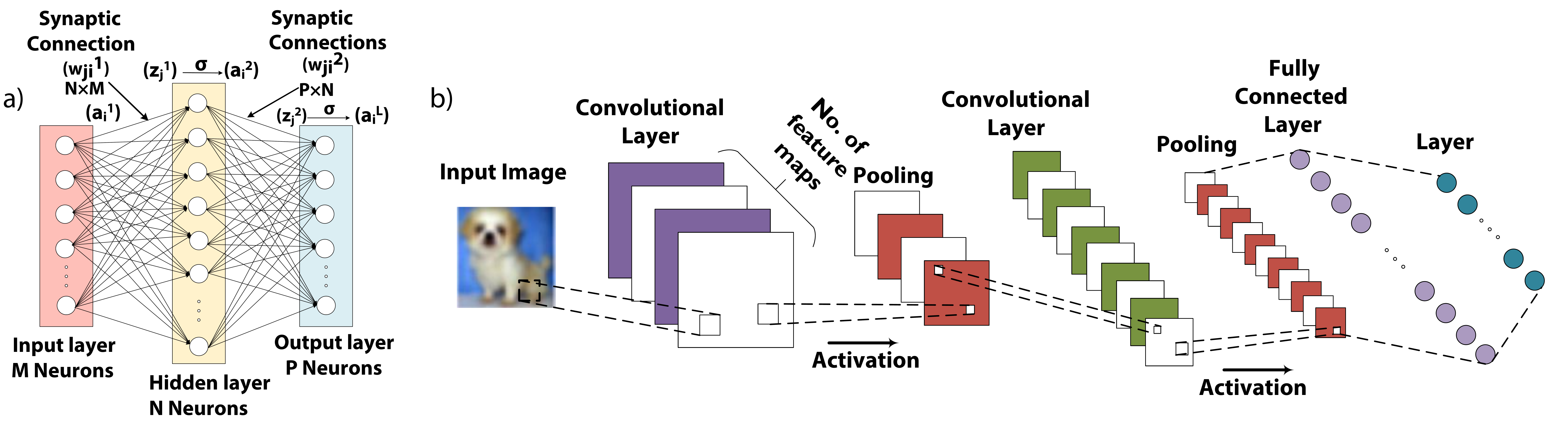}
		\vspace{-4mm}
		\caption{(a) Fully connected 3-layered neural network showing the input layer, hidden layer, and an output layer. Each neuron in a particular layer is fed by weighted sum of all inputs of the previous layer and it performs a sigmoid operation on the sum to provide the inputs for the next layer. (b) CNN Architecture with different convolutional and pooling layers terminated by a fully connected layer.}
		\vspace{-4mm}
		\label{fig:fcn}
\end{figure*}
In this work, we analyze the impact of non-idealities such as source resistance, neuron resistances, and synaptic weight variations in hardware implementations of neuromorphic crossbars. We show how such non-idealities can significantly degrade the accuracy when traditional training methodologies are employed. The presence of these parasitic elements also severely limits the crossbar sizes. As a solution, we propose an ex-situ technology aware training algorithm that mathematically models the aforementioned non-idealities and accounts for the same in the traditional backpropagation algorithm. Such a technique not only preserves the accuracy of an ideal network appreciably but also allows us to use larger crossbar sizes without significant accuracy degradation. The key highlights of our work are as follows:

\begin{enumerate}
\item We mathematically model the effect of source resistance, neuron resistance, and variations in synaptic conductance on the output currents of a neuromorphic crossbar. We establish the validity of our model by comparing against SPICE-like simulations of resistive networks. 
\item We analyze the impact of these non-idealities on the accuracy of two types of image recognition tasks with varying amounts of non-ideality within relevant technological limits.
\item We propose a training algorithm which incorporates the mathematical models of the crossbar non-idealities and modifies the standard training algorithm in an effort to restore the ideal accuracy. 
\end{enumerate}

\section{Crossbar Implementation of Neural Networks}
\subsection{Types of network topologies}
\subsubsection{Fully Connected Networks}
Traditionally, deep neural networks such as deep belief nets (DBNs) comprise of multiple layers of interconnected units. Fully connected networks (FCN) involve a series of neuron layers between the input and the output layers. The output of each neuron in a layer is connected to the inputs of all the neurons in the subsequent layer. Fig. \ref{fig:fcn}(a) shows a 3-layered fully connected network consisting of a single hidden layer between the input and output layers. 
\subsubsection{Convolutional Networks} Complex image recognition datasets comprise of objectively different classes where global weight mapping like FCNs prove to be less efficient. As an alternative, convolutional neural networks (CNN) have been recognized as a more powerful tool for complex image recognition problems using locally shared weights to learn common spatially local features. As shown in Fig. \ref{fig:fcn}(b), CNNs consist of several layers performing operations like convolution, activation, and pooling, finally terminating with a fully connected layer. During the convolution operation, each filter bank, or kernel, is slid across the input to that layer to obtain a dot product between the input and the weights, known as the feature map. The number of output maps of each convolutional layer denotes the number of different feature maps learned in that layer.  Thus a convolution operation captures the spatially local features of an input image. Convolution of a $m\times m$ input map with kernel of size $n\times n$ yields an output map of size $((m-n+2p)/s+1)\times((m-n+2p)/s+1)$, where $s$ is the stride of the filter and $p$ is the padding. In practice, $s$ and $p$ are chosen such that the original input size is preserved. The activation layer which can be ‘RELU’ \cite{nair2010rectified}, ‘sigmoid’\cite{hopfield1984neurons}, or other non-linear functions, introduces a non-linearity in the network\cite{glorot2010understanding}. The pooling layer reduces the dimensionality of the output map. Most commonly used pooling techniques are average and max-pooling\cite{jarrett2009best}. Finally, the fully connected layer uses the learned features to classify the images. In essence, a fully connected layer could also be represented by a convolutional layer where the kernel size is equal to the input size. 
\begin{figure}[t]
		\centering
		\includegraphics[width=2.1in,keepaspectratio]{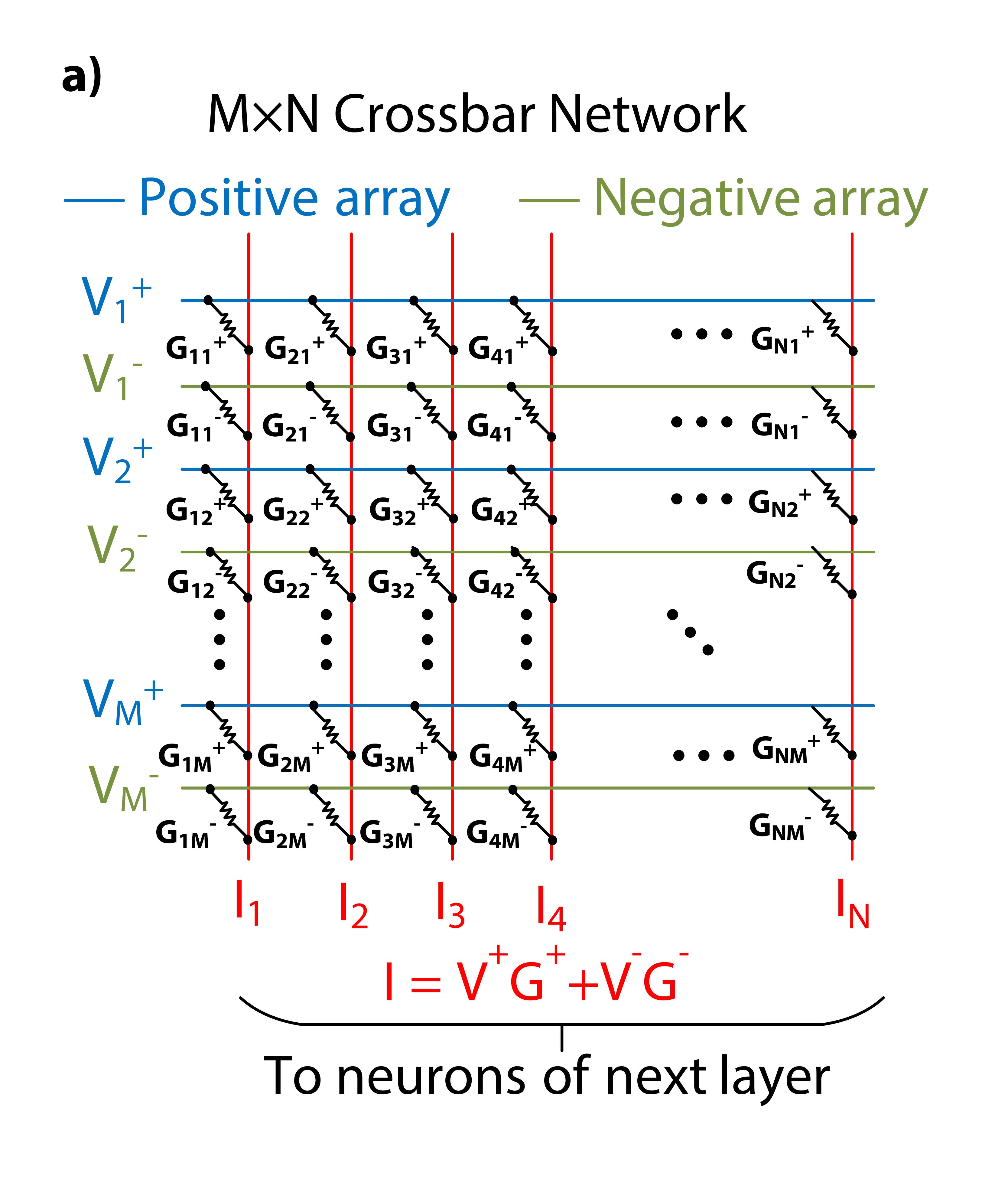}
      	\includegraphics[width=3.4in,keepaspectratio]{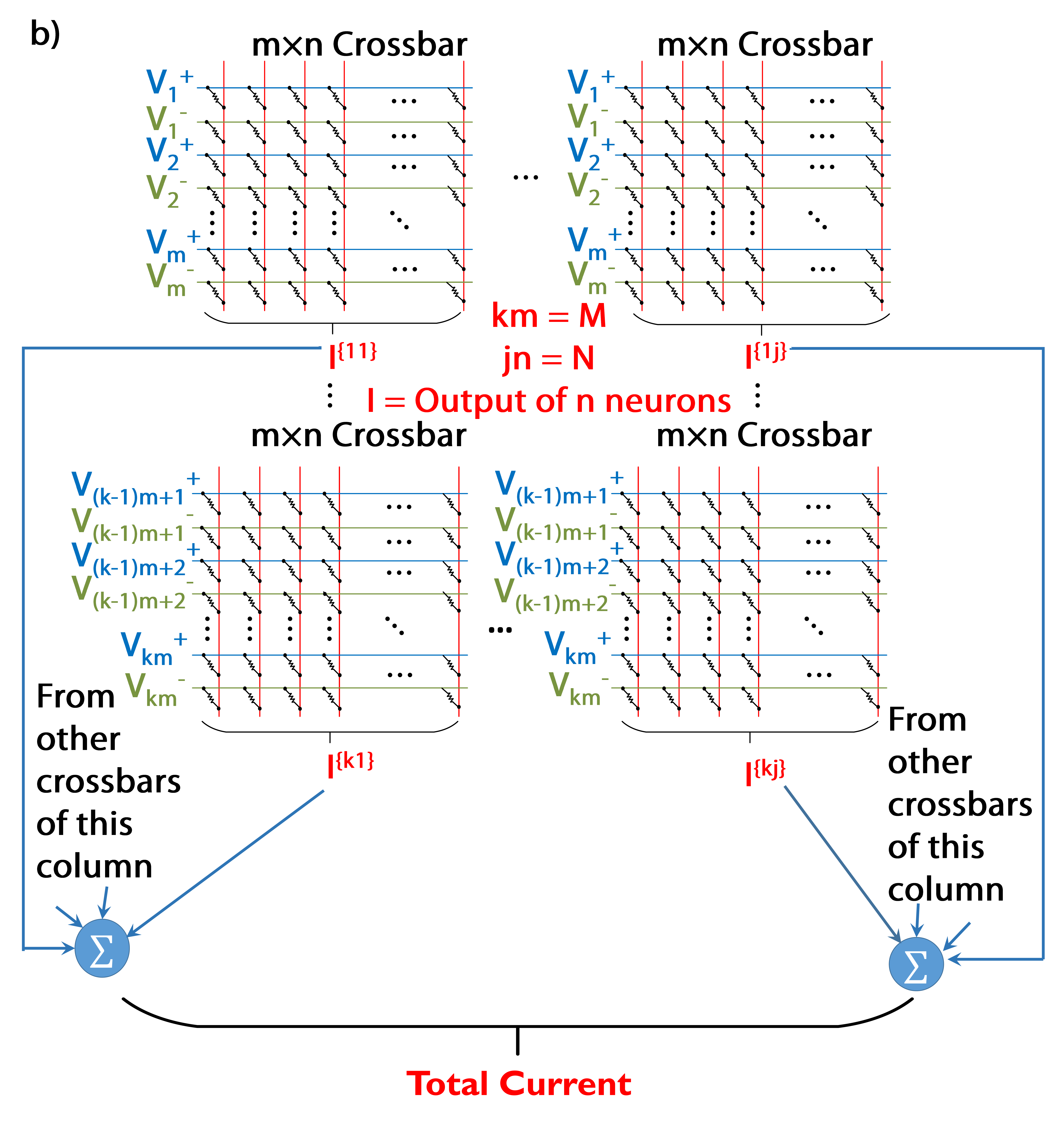}
		\vspace{-4mm}
		\caption{(a) Hardware implementation of a single fully connected network layer represented by two resistive crossbar arrays. The output of the crossbar will be fed to another crossbar representing the next layer. (b) An arrangement of multiple sub-crossbars to realize the functionality of a large crossbar.}
		\vspace{-4mm}
		\label{fig:cbrep}
\end{figure}
\begin{table*}[t]
\centering
\begin{threeparttable}
\caption{Resistance Ranges for Various technologies}
\label{tech}
\begin{tabular}{|l|l|l|l|l|}
\hline
Technology                                  & ($R_{on}$, $R_{off}$)                               & Considered Range ($R_{low}$, $R_{high}$) & $R_s/R_{high}(\%)$ & $R_{neu}/R_{high} (\%)$ \\ \hline\hline
TiO\textsubscript{2}\cite{Berdan_2016} & 15k, 2M                                             & 40k,600k                                 & 0.033 - 0.13        & 0 - 0.033               \\
Ag/Si \cite{Kim_2012}                    & 25k,10M                                             & 100k,1.5M                                & 0.013 - 0.053         & 0 - 0.013                 \\
TaOx \cite{Kim_2016}                     & 1k,1M                                               & 20k,300k                                 & 0.067 - 0.27       & 0 - 0.067               \\
Spintronics*                                & Function of MTJ oxide thickness \cite{Fong_2011} & 40k, 400k                                & 0.05 - 0.2        & 0-0.05                  \\
PCM \cite{Eryilmaz_2014}                 & 10k, 3M                                             & 60k, 900k                                & 0.022 - 0.08      & 0 - 0.022 \\ \hline

\end{tabular}
\begin{tablenotes}
\item *The spintronic analysis is done based on predictive measures of $R_{off}/R_{on}$\cite{Hirohata_2015}
\end{tablenotes}
\begin{tablenotes}
\item $R_s$ range - 200 to 800 $\Omega$, $R_{neu}$ range - 0 to 200 $\Omega$
\end{tablenotes}
\end{threeparttable}
\end{table*}
\subsection{Hardware representations of Neural networks}
In hardware realizations of neural networks, the synaptic connections between the neurons of two adjacent layers are represented using a resistive crossbar. The weights are represented in terms of conductance and the inputs are encoded as voltages. Convolutional layers have locally concentrated connections, hence each filter bank is represented by a crossbar of equivalent size. The input to the crossbar is a subset of the image being sampled by the kernel. Each element of the output map is calculated through time multiplexing of the outputs from a particular crossbar for different subsets of the image. This is repeated for each filter bank to obtain different output maps. In contrast, fully connected layers have all possible connections between input and the output and the entire connection matrix can be represented by a crossbar. The basic computational function of any layer is a dot product and can be seamlessly performed by representing the weights as the resistances in a crossbar fashion. The output current of $j^{th}$ neuron of each crossbar is computed as $I_j = \sum{V_i^{+}G_{ji}^{+}+V_i^{-}G_{ji}^{-}}$, where $V_i$ is the input voltage corresponding to $i^{th}$ input neuron and $G_{ji}$ represents the conductance corresponding to the synaptic weights between the neurons. Two resistive arrays are deployed to account for bipolar weights. The input to the positive array is $+V_i$ whereas the input to the negative array is $-V_i$. The weight matrix [$w_{ji}$] is mapped to a corresponding conductance range ($G_{low}$, $G_{high}$) $\subset$ ($G_{on}$, $G_{off}$). To represent bipolar weights, the conductance of the synapse connecting the $j^{th}$ neuron in the next layer to the $i^{th}$ input is denoted by a positive ($G_{ji}^{+}$) component and a negative ($G_{ji}^{-}$) component. For positive (negative) weights, the programming is done such that $G_{ji}^{+}(G_{ji}^{-}) = |w_{ji}|G_{high}$ and $G_{ji}^{-}(G_{ji}^{+}) = 0$ (no connection). Fig. \ref{fig:cbrep}(a) shows a crossbar implementation of a fully connected neural network. \par

As mentioned earlier, crossbar arrays could suffer from non-ideal effects and incur limitations on their sizes. As a result, larger crossbars are divided into smaller crossbars and the output of each crossbar is time-multiplexed to obtain the desired functionality of the entire crossbar. Fig. \ref{fig:cbrep}(b) shows how multiple small crossbars can be efficiently mapped to realize the functionality of a large crossbar in a particular layer. The small size of the crossbar reduces fan-out and fan-in, thus minimizing the impact of non-idealities. FCNs, being densely connected, are severely affected by hardware imperfections, especially when implemented on large crossbars. Convolutional layers in CNNs are usually implemented on very small crossbars and are thus insensitive to non-ideal effects. However, the final fully connected layers which acts as a classifier can be significantly affected by these non-idealities due to their large sizes. In this work, we are thus considering the impact of non-idealities on FCNs, and fully connected layers of CNNs.  
\subsection{Training}
The training of Artificial Neural Networks (ANN) are traditionally done off-chip through the standard backpropagation algorithm which updates weight matrices using gradient descent technique\cite{RUMELHART_1988}. It is important to note down the vital aspects of the algorithm here in relevance to the later sections. The basic algorithm updates weights based on the gradients of a cost function. The cost function depends on the error computed from the feed-forward network which assumes a form : $C = \frac{1}{2}\sum{(y_j - a_j)^2}$, where $y_j$ is the expected output and $a_j$ is actual output from the $j^{th}$ neuron in the output layer. The sensitivity of the errors for each layer are calculated from the derivatives of the cost function with respect to the outputs and weights and after each iteration, the weights are updated based on those of the corresponding layer. The detailed description of the algorithm is well documented\cite{RUMELHART_1988}. In this work, we focus on the aspects of the algorithm pertinent to fully connected layers and we build mathematical models to account for the non-idealities experienced by the hardware implementation of neuromorphic crossbars.

\subsection{Technologies}
Various technologies have been explored for crossbar implementations of neural networks. Memristive crossbars based on different material systems (like $TaO_x$\cite{Wang_2014}, $TiO_2$\cite{Alibart_2013}, $Ag/Si$\cite{Jo_2010} etc) have been proposed to realize neuromorphic functionality in an energy efficient manner. Phase change materials (PCM)\cite{Eryilmaz_2014} have also been investigated as potential candidates for neuromorphic computing due to their high scalability. More recently, neurons and synapses implemented with spintronic devices\cite{Sengupta_2016,Sengupta2_2016} have shown great promise in performing ultra-low power neuromorphic computing. However, each technology suffers from specific drawbacks. An important metric in regard of resistive crossbars for neuromorphic systems is the the ratio of the high resistance state ($R_{off}$) and the low resistance state ($R_{on}$) of the synaptic device. Usually, a high $R_{off}/R_{on}$ ratio is desired for a near-ideal implementation of the weights in a neuromorphic crossbar. Moreover, in the light of non-ideal systems, higher values of $R_{on}, R_{off}$ may be less significantly impacted by parasitic resistances. In this work, we have chosen a maximum to minimum conductance ($G_{low} = \alpha/R_{off}$, $G_{high} = 15G_{low}$, $\alpha$ is a parameter of choice) ratio of 15 which is a potentially realizable predictive measure for all memory technologies\cite{Hirohata_2015,Borghetti_2010,Eryilmaz_2014}.


\section{Modeling the non-idealities}
\begin{figure}[h!]
		\centering
		\includegraphics[width=3.2in,keepaspectratio]{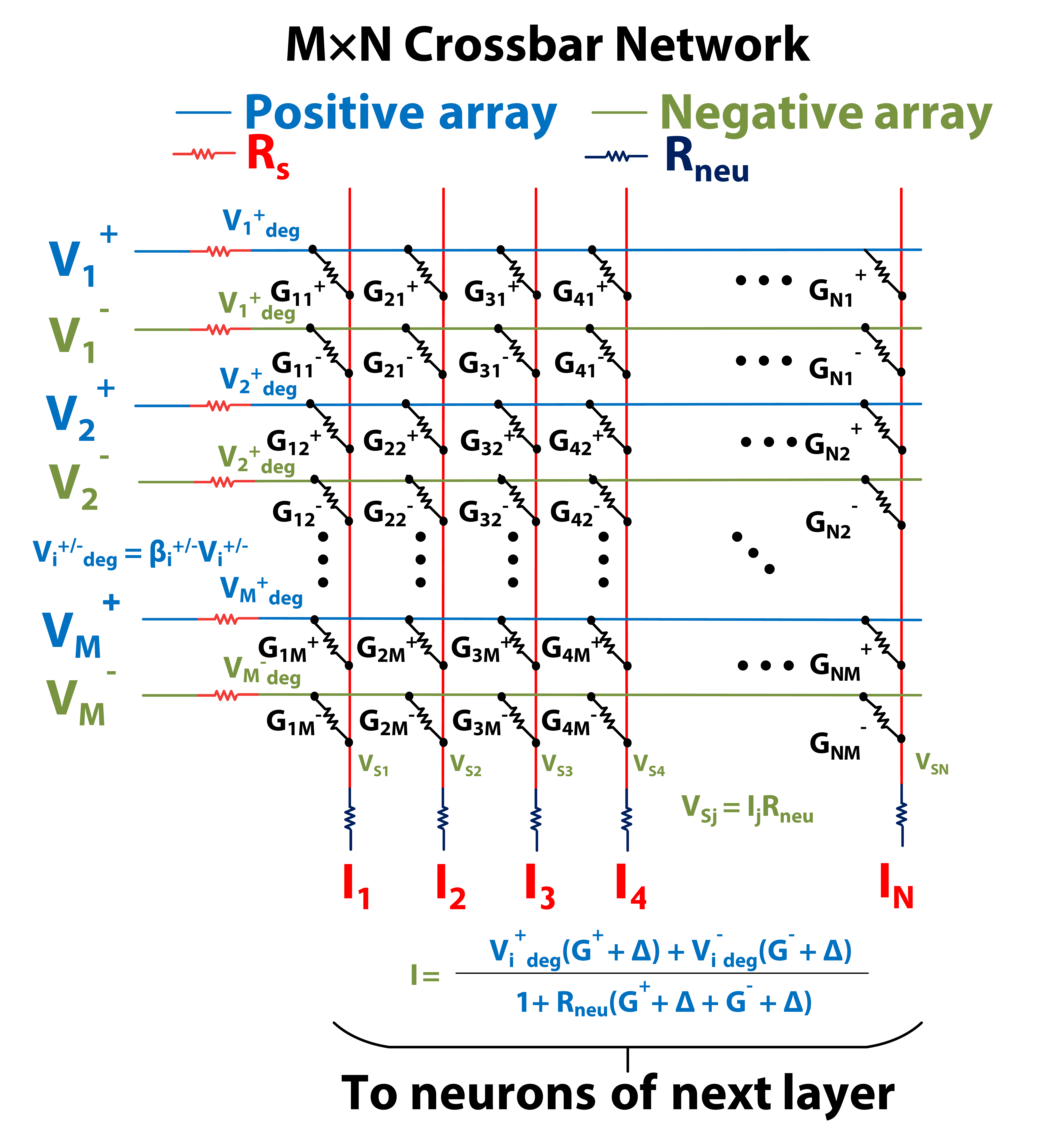}
		\caption{Crossbar Architecture showing non-ideal elements like source and neuron resistances. The final output current equation is modified by the impact of these non-ideal elements.}
		\vspace{-2mm}
		\label{fig:NICB}
\end{figure}
Memristor based neuromorphic crossbar designs leverages its inherent capability of matrix multiplication to provide high accuracy at a relatively modest computational cost\cite{hu2016dot}. However, the memristor technology is still in its nascent stage. Thus, the hardware implementation of such crossbars may suffer various kinds of non-ideal effects arising from memristor device variations, parasitic resistances as well as non-idealities in sources, and sensing neurons. 
In this work, we have considered three kinds of non-idealities that arise in crossbar implementations, namely,
\begin{enumerate}
\item Neuron Resistance ($R_{neu}$)
\item Source Resistance ($R_{s}$)
\item Memristive resistance variations
\end{enumerate}
To perform an analysis of the impact non-idealities might have on accuracy of recognition task, it is important to note the ratio of non-ideal resistances to the synaptic resistances for a particular technology. Table. \ref{tech} shows the range of considered resistance ratios to synaptic resistances $R_s/R_{high}$ and $R_{neu}/R_{high}$ for various technologies, considering relevant values of source ($R_{s}$) and neuron ($R_{neu}$) resistances. 
\subsection{Neuron Resistance}
The resistance offered by the neuron in a neuromorphic crossbar varies from technology to technology. In many cases, such as, PCM technology, the resistance of the neuron is not a hardware issue as the crossbar outputs are sensed through a sense amplifier, where virtual ground at the input eliminates the voltage drop across the neuron. However, in spintronic crossbars\cite{Sengupta_2016}, crossbar outputs are fed to the neuron as a current stimulus and thus, the resistance of the neuronal device becomes relevant. Fig. \ref{fig:NICB} shows the effect of neuron resistance on the crossbar output. This can be mathematically modeled to modify Eqn (1) as:
\begin{equation}
I_j = \frac{\sum{V_i^{+}G_{ji}^{+}+V_i^{-}G_{ji}^{-}}}{1+R_{neu}\sum{G_{ji}^{+}+G_{ji}^{-}}}
\end{equation}
Here, $I_j$, $V_i^{+/-}$, $G_{ji}^{+/-}$ and $R_{neu}$ carry the same meaning as described in earlier sections. 
Eqn (1) can be derived by applying Kirchoff's law at the output nodes of the crossbar and considering the voltage drop across the neuron to be $V_{j,neu} = I_j\times R_{neu}$. It is evident that the denominator is close to 1 for smaller arrays as $G_{ji}$s are much smaller than neuron conductances (resistances of the order of a few hundred ohms \cite{Sengupta_2016}). However, larger arrays could lead to $G_{neu} = 1/R_{neu}$ being comparable to sum of the conductances in a particular column. More specifically, a higher number of rows in the crossbar lead to enhanced impact of neuron resistance.  
\subsection{Source Resistance}
The source resistance ($R_s$) in a neuromorphic crossbar could arise due to non-ideal voltage sources and input access selectors lumped together. The input voltages to crossbar gets degraded due to $R_s$ and the degradation can be mathematically modeled as:
\begin{align}
V_{i,deg}^+ = V_i^+\frac{1/R_s}{1/R_s+\sum{\frac{1}{R_{ij}^++R_{neu}}}}\\
V_{i,deg}^- = V_i^-\frac{1/R_s}{1/R_s+\sum{\frac{1}{R_{ij}^-+R_{neu}}}}
\end{align}
Here $R_{ij}^{+/-}$ is the resistance of the synaptic element between the $i^{th}$ row and $j^{th}$ column in the positive or negative array.
The model ignores the effect of sneak paths. In neuromorphic crossbars, all the inputs are simultaneously active. As the IR drops in the metal lines are negligible, all the nodes in a particular row are supplied by the degraded source voltage of that row. As all the rows are supplied by voltages of same polarity, even the shortest possible current sneak path will experience a low potential difference. Thus, the current through the series connection of the synaptic memristor and neuron would be primarily dependent on the degraded supply voltage and effective series resistance. We have verified the validity of the model by comparing against SPICE-like simulations, which is described in more detail in Section IV B.
\subsection{Memristive Conductance Variations}
The weights obtained from the training algorithm are usually discretized in order to be represented as memristive synapses. In this work, we have used a 4-bit discretization technique where we have used a $R_{high}/R_{low}$ ratio of 15, relevant to the technologies considered. We have mapped the weights such that the maximum weight always maintains the $R_{high}/R_{low}$ ratio to the minimum weight. We have chosen the maximum and minimum weight limits so as to minimize the accuracy degradation due to discretization. To analyze the impact of chip-to-chip variation of weights, we have introduced weight variations in terms of standard deviation ($\sigma$) errors, ranging from -2$\sigma$ to +2$\sigma$ after discretization. This implies that all the memristive devices on a neuromorphic chip suffer the same variation at a particular process corner. The weight variations are incorporated in the mathematical model as a $\Delta$ variation to the conductances. 
\subsection{Proposed Training Algorithm}
The mathematical representations of the non-idealities are finally collated and incorporated in the feed-forward path and the backpropagation algorithm for training the ANN. Weights $w_{ji}$ and inputs $a_{i}$ replaces the conductances $G_{ji}$ and voltages $V_{i}$ respectively in Eqn (1) and Eqn (2). The symbol $z_j$ is used to represent the current output of the crossbars $I_j$ corresponding to $j^{th}$ neuron of the next layer. We assume that the neuronal function receives a current input and provides a voltage output. For the sake of simplicity, we assume ideal mathematical representations of activation functions like ‘RELU’ \cite{nair2010rectified} and ‘sigmoid’\cite{hopfield1984neurons}. As described in Section. II A, the ideal crossbar output of the $j^{th}$ column in any layer is given by $z_j = \sum\limits_i{a_i\times w_{ji}}$. The modified crossbar output can be computed as follows:
\begin{gather}
z_j^{l} = \frac{\sum{a_{i,deg}^+w_{ji,vary}^{+}+a_{i,deg}^-w_{ji,vary}^{-}}}{\gamma_j}\\
\gamma_j = 1+R_{neu}\sum\limits_i{w_{ji,vary}^++w_{ji,vary}^-}\nonumber
\end{gather}
where, 
\vspace{-8mm}
\begin{gather*}
a_{i,deg}^+ = a_i\frac{1/R_s}{\beta_{i}^+}\\
a_{i,deg}^- = -a_i\frac{1/R_s}{\beta_{i}^-}\\
w_{ij,vary}^{+/-} = w_{ij}^{+/-}+\Delta\\
\beta_{i}^{+/-} = 1/R_s+\sum{\frac{1}{R_{ij}^{+/-}+R_{neu}}}\\
R_{ij}^{+/-} = 1/w_{ij,vary}^{+/-}
\end{gather*}
As described earlier, two weight matrices are deployed to account for bipolar weights in the original weight matrix $W = [w_{ji}]$. Positive (Negative) inputs are fed to the positive (negative) weight array. The weight matrices are created such that $w_{ji}^{+}(w_{ji}^{-}) = 0$ for all i,j for which $W_{ji}<0( > 0) $ and $w_{ji}^{+}(w_{ji}^{-}) = W_{ji}$ for all i,j for which $W_{ji}>0(<0)$. Note that mapping the weights to a particular conductance range is equivalent to multiplication by a scaling factor as we have already discretized the weights based on a maximum to minimum weight ratio equal to $G_{high}/G_{low} = 15$. Thus an equivalent representation in terms of conductance would be $G_{ji}^{+/-} = W_{ji}G_{high}$. \par
The output of each crossbar is passed as inputs to the next crossbar through a sigmoid function such that $a_i^{L+1} = \sigma (z_i^{L})$ (where L is the layer index). The backpropagation algorithm is modified to account for the modified crossbar functionality. As described earlier, learning in neural networks relies on computation of gradients of a cost function. Here, it is calculated from the error between the expected and the actual output of the output layer neurons in the form of $C = \frac{1}{2}\sum{(y_j - a_j^L)^2}$. The delta-rule in the backpropagation algorithm \cite{hecht1988theory} involves calculation of $\delta$ for each layer accounting for the change in the cost function for unit change in inputs to that particular layer. Thus, $\delta$ for layer $l$ can be written as:\par
For output layer,
\begin{gather}
\delta_j^L = \frac{\partial C}{\partial z_j^L} = \sum{\frac{\partial C}{\partial a_j^{L}}\frac{\partial a_j^{L}}{\partial z_j^{L}}} = (a_j^L - y_j)\sigma^{'}(a_j^L)\\
\intertext{For other layers,}
\delta_j^l = \frac{\partial C}{\partial z_j^l} = \sum\limits_{k}{\frac{\partial C}{\partial z_k^{l+1}}\frac{\partial z_k^{l+1}}{\partial z_j^{l}}} = \sum\limits_{k}{\delta_k^{l+1}\frac{\partial z_k^{l+1}}{\partial z_j^{l}}}\\
\frac{\partial z_k^{l+1}}{\partial z_j^{l}} = \frac{\partial z_k^{l+1}}{\partial a_j^{l}}\frac{\partial a_j^{l}}{\partial z_j^{l}} = \frac{\partial z_k^{l+1}}{\partial a_j^{l}}\sigma^{'}(a_j^l)\nonumber\\
\frac{\partial z_k^{l+1}}{\partial a_j^{l}} = \frac{\frac{a_{j,deg}^{+,l}}{a_j^l}w_{jk,vary}^{+} - \frac{a_{j,deg}^{-,l}}{a_j^l}w_{jk,vary}^{-}}{\gamma_j}
\end{gather}
Finally, the $\delta$s of each layer are used to compute the weight updates as:
\begin{gather}
dw_{jk}^l = \frac{\partial C}{\partial w_{jk}^l} = \frac{\partial C}{\partial z_{j}^l}\frac{\partial z_j^l}{\partial w_{jk}^l} = \delta_j^l\frac{\partial z_j^l}{\partial w_{jk}^l}\\
\frac{\partial z_j^l}{\partial w_{jk}^l} =\frac{\gamma_{j}(a_{k,deg}^+(1-\frac{w_{kj}^+}{\beta_k^+})+a_{k,deg}^-(1-\frac{w_{kj}^-}{\beta_k^-}))-R_{neu}z_j^l\gamma_j}{\gamma_j^2}
\end{gather}
To simulate the impact of non-idealities on varying crossbar size, we divide the large crossbars of size $M\times N$ into several smaller crossbars of size $m \times n$. Fig. \ref{fig:fcn}(c)  shows the network architecture of combining smaller crossbars to realize the neuromorphic functionality of larger crossbars. The source degradation factor $\beta_i$ is more prominent for larger number of columns  as it depends on the term $\sum\limits_j{1/(R_{ij}+R_{neu})}$ summed over the columns. The neuron resistance degradation factor $\gamma_j$, on the other hand, increases with the number of rows due to its dependence on the term $\sum\limits_i{w_{ji}}$, summed over the rows. Thus, the combined effect of these two non-idealities is expected to have a higher impact on the network for larger crossbars. 
\section{Simulation Framework}
\subsection{Model simulations}
The model described in the previous section was implemented on FCNs using the MATLAB\textsuperscript{\textregistered} Deep Learning Toolbox\cite{palm2012prediction} and CNNs using MatConvNet \cite{Vedaldi_2015}. 
\subsubsection{FCN}
A 3-layered neural network was employed to recognize digits from the MNIST Dataset. The training set consists of 60000 images, while the testing set consists of 10000 images. The input layer consists of 784 neurons designated to carry the information of each pixel of each 28$\times$28 image. The hidden layer consists of 500 neurons and the output layer has 10 neurons to recognize 10 digits. The neuron transfer function was chosen to be the sigmoid function which can be written as $\sigma(x) = \frac{1}{1+e^-x}$. 
\subsubsection{CNN}
\begin{table}[t]
\centering
\caption{CNN Architecture}
\label{my-label}
\begin{tabular}{|c|}
\hline
Input 32 $\times$ 32 RGB image    \\ \hline
5 $\times$ 5 conv. 64 RELU        \\
2 $\times$ 2 max-pooling stride 2 \\ \hline
5 $\times$ 5 conv. 128 RELU       \\
2 $\times$ 2 max-pooling stride2  \\ \hline
3 $\times$ 3 conv. 256 RELU       \\
2 $\times$ 2 avg-pooling stride 2 \\ \hline
4 $\times$ 4 conv. 512 Sigmoid (fully connected)    \\
0.5 Dropout                       \\ \hline
1 $\times$ 1 conv. 10  (fully connected)           \\ \hline
10-way softmax                    \\ \hline
\end{tabular}
\end{table}
For the classification of more complex dataset CIFAR-10, we have used a network with RELU-activated convolutional layers and a sigmoid-activated fully connected layer. The architecture is represented as 32$\times$32$\times$3-64c5-2s-128c5-2s-256c3-2s-512o-10o.The details of the layers are provided in Table. II. Each convolutional layer is followed by a batch-normalization layer for better performance. We concentrate our analysis on the fully connected layers of the network as the initial convolutional layers possess local connections implemented on small crossbars equal to the kernel sizes.
\subsection{SPICE-like Simulations for validation}
\begin{figure}[h!]
		\centering
		\includegraphics[width=3.2in,keepaspectratio]{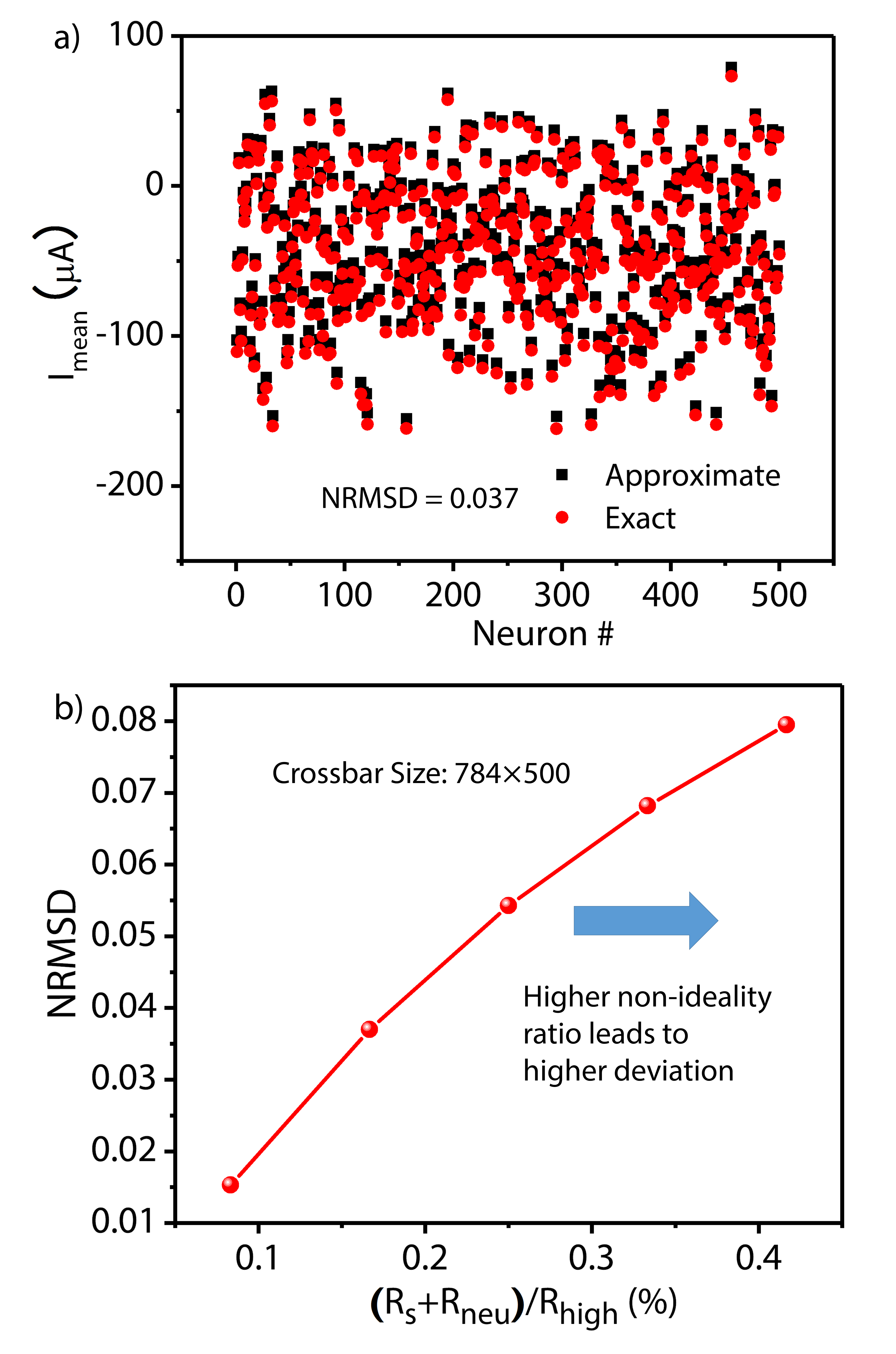}
		\caption{(a) Distribution of output currents ($I_{mean}$), averaged over 100 images, across 500 neurons in the hidden layer comparing the approximate model to SPICE-like simulation framework. (b) Variation of Normalized Root Mean Square Deviation (NRMSD) with non-ideality ratio. NRMSD is close to zero for the relevant range of non-idealities.}
		\vspace{-4mm}
		\label{fig:comp}
\end{figure}

Each fully connected layer for both FCNs and CNNs can be implemented in a crossbar architecture comprising of all possible connections. A SPICE-like framework was implemented in MATLAB\textsuperscript{\textregistered} by creating a netlist of all connections, voltage source, source and neuron resistances in such resistive crossbars and evaluating the voltages at each node by solving the conductance matrix: $[V] = [G]^{-1}[I]$. The framework was benchmarked with HSPICE\textsuperscript\textregistered. This framework was used to calculate the output of non-ideal crossbars on application of the inputs from the MNIST dataset as voltages. The resistances of the crossbar elements $R_{ji}$ were determined such that $R_{ji} = 1/w_{ji}$, where $w_{ji}$ are the weights determined by the ideal training scheme described in the previous subsection. The output obtained by showing 100 images of the testing set was averaged and the distribution was compared with the mathematical model simulations. Fig. \ref{fig:comp}(a) shows the comparison in the distribution of output currents of a crossbar where the approximate model shows good agreement with the exact SPICE-like simulations. Fig. \ref{fig:comp}(b) shows that the normalized root mean square deviation (NRMSD) between the two techniques for various $(R_s+R_{neu})/R_{high}$ combinations remains very close to zero for relevant values. It is to be noted that the validation of our approximate model was important in the context of reducing the training time as the matrix operations could be more efficiently performed using the mathematical model than simulating the network for each input image in HSPICE\textsuperscript\textregistered. 
\section{Results and Discussion}
We analyzed the impact of technological constraints in crossbar implementations on both FCNs and CNNs. As fully connected layers form the crux of classification in both network topologies, it is expected that such non-ideal conditions will have similar detrimental effects on both. We present the detailed impact of each non-ideality on FCNs and CNNs for better understanding.
\par
We consider a 3-layered FCN and a CNN architecture described in Table. \ref{my-label} to analyze the impact of the non-idealities on the accuracy of recognition task on MNIST and CIFAR-10 datasets respectively. The other convolutional layers in the CNN are usually implemented using small crossbars and hence do not suffer significant effects of non-ideal resistances. \par
First, the neural networks were trained under ideal conditions using the training set. Then, the non-ideal model was included in the feed-forward path and the ideally trained network was tested using the testing set to determine the performance degradation due to the non-idealities. Next, the technology aware training algorithm was implemented by incorporating the mathematical formulation of the non-idealities in the standard training iterations of feed-forward and backpropagation as described in the Section III D. For each iteration, the weights were discretized as described in Section III C. The testing accuracy of an ideally trained FCN with a sigmoid neuronal function was 98.12\% on MNIST and that of an ideally trained CNN was 85.6\% on CIFAR-10 datasets. The accuracy degradations discussed in this section has been calculated with respect to these ideal testing accuracies such that Accuracy Degradation (\%) = Ideal Accuracy (\%) - Accuracy Obtained (\%).
\begin{figure}[t]
		\centering
		\includegraphics[width=3.1in,keepaspectratio]{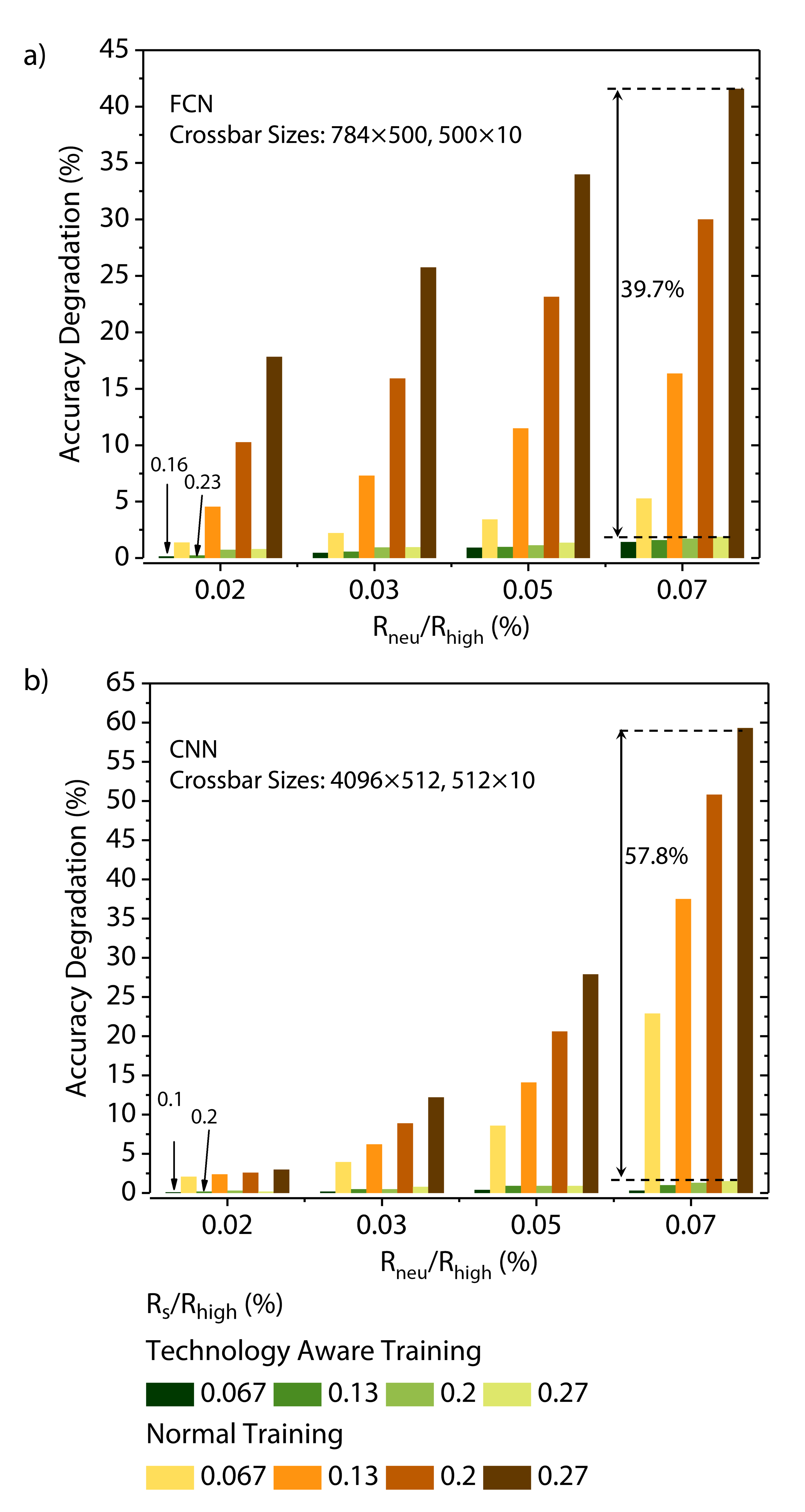}
		\vspace{-4mm}
		\caption{Accuracy degradation v/s varying $R_{neu}/R_{high}$ ratio for different $R_{s}/R_{high}$  combinations comparing technology aware training scheme with normal training for (a) FCN and (b) CNN.}
		\vspace{-4mm}
		\label{fig:neuronvar}
\end{figure}
We use the parameters $R_s/R_{high}$ and $R_{neu}/R_{high}$ to denote the ratios of the non-ideal resistances and the maximum synaptic resistance. 
\begin{figure}[t]
		\centering
		\includegraphics[width=3.1in,keepaspectratio]{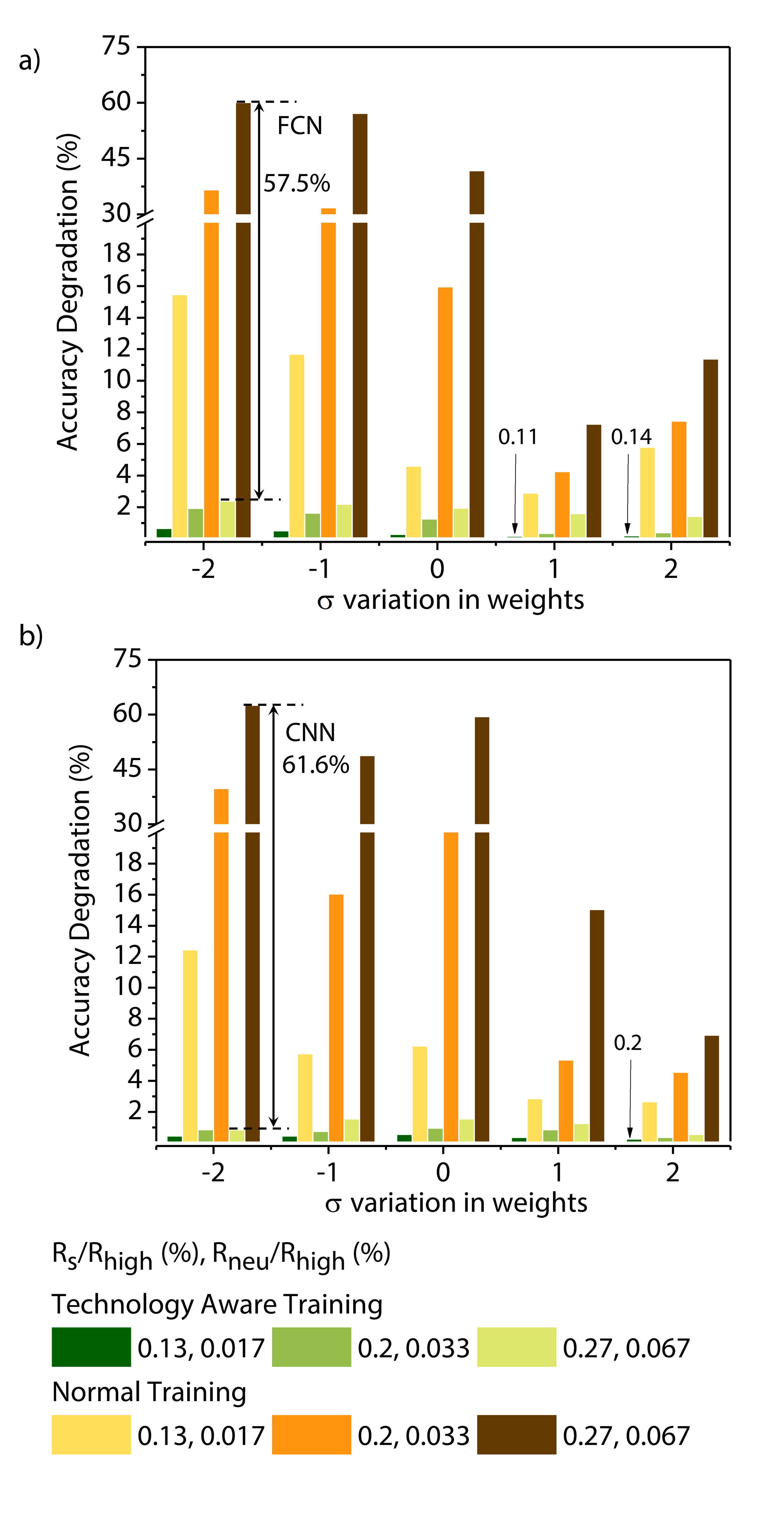}
		\vspace{-4mm}
		\caption{Accuracy degradation v/s $\sigma$ variations in weights for various $R_s/R_{high}$ and $R_{neu}/R_{high}$ combinations comparing the technology aware training scheme with normal training for (a) FCN and (b) CNN.}
		\vspace{-6mm}
		\label{fig:wvar}
\end{figure}
\subsubsection{Source and Neuron Resistance}
Fig. \ref{fig:neuronvar}(a) and \ref{fig:neuronvar}(b) shows the accuracy degradation for different $R_{neu}/R_{high}$ and $R_{s}/R_{high}$ combinations in FCN and CNN, respectively. The effect of the non-ideal resistances on the performance of the network predictably worsens monotonically with higher $R_s/R_{high}$ and $R_{neu}/R_{high}$ ratios. It can be observed that with normal training methods, the non-ideal resistances result in accuracy degradation for FCN: up to 41.58\% for $R_{neu}/R_{high} = 0.07\%$ and $R_{s}/R_{high} = 0.27\%$. Our proposed training scheme incorporates the impact of non-idealities and achieves significant restoration of accuracy, within 1.9\% of the ideal accuracy, for the worst case combination of resistances considered, shown in Fig. \ref{fig:neuronvar}(a). \par
In case of CNNs, we show that due to the large crossbar sizes of the fully connected layers in the CNN, it can suffer up to 59.3\% degradation in accuracy for the worst case non-ideal resistances considered. Our proposed algorithm, on the other hand, achieves an accuracy within 1.5\% of the ideal accuracy (Fig. \ref{fig:neuronvar}(b)), considering the largest crossbar sizes for the architecture. 

\subsubsection{Weight variations}
On-chip crossbar implementations suffer from chip-to-chip device variations. To account for such variations, we form a defect weight matrix, and include it in the feed-forward network, as described in detail in Section III C. We have considered up to $\pm$2$\sigma$ variation in the synaptic weights. Fig. \ref{fig:wvar} shows the impact of such device variations on the accuracy of FCN and CNN for different combinations of $R_{s}/R_{high}$ and $R_{neu}/R_{high}$. Predictably, changes in the positive direction reduces the accuracy degradation from the nominal (no variation) case as it enhances the significance of the neurons. However, changes in the negative direction slightly degrades the accuracy from the nominal case. It is observed that a $-2\sigma$ variation can result in an accuracy degradation of up to 59.9\% for $R_{neu}/R_{high} = 0.067\%$ and $R_{s}/R_{high} = 0.27\%$ in FCN. By accounting for these variations in the backpropagation algorithm, our proposed training methodology successfully restores the accuracy within 2.34\% of the ideal accuracy for worst case of non-idealities considered, as shown in Fig. \ref{fig:wvar}(a). \par
Weight variations in the negative direction also adversely affect CNNs where $-$2$\sigma$ variation can result in an accuracy degradation of 62.4\% considering the non-ideal resistances mentioned above. Our proposed algorithm achieves an accuracy within 0.8\% of the ideal testing accuracy as shown in Fig. \ref{fig:wvar}(b). 

\subsubsection{Crossbar Size}
\begin{figure}[t]
		\centering
		\includegraphics[width=3.1in,keepaspectratio]{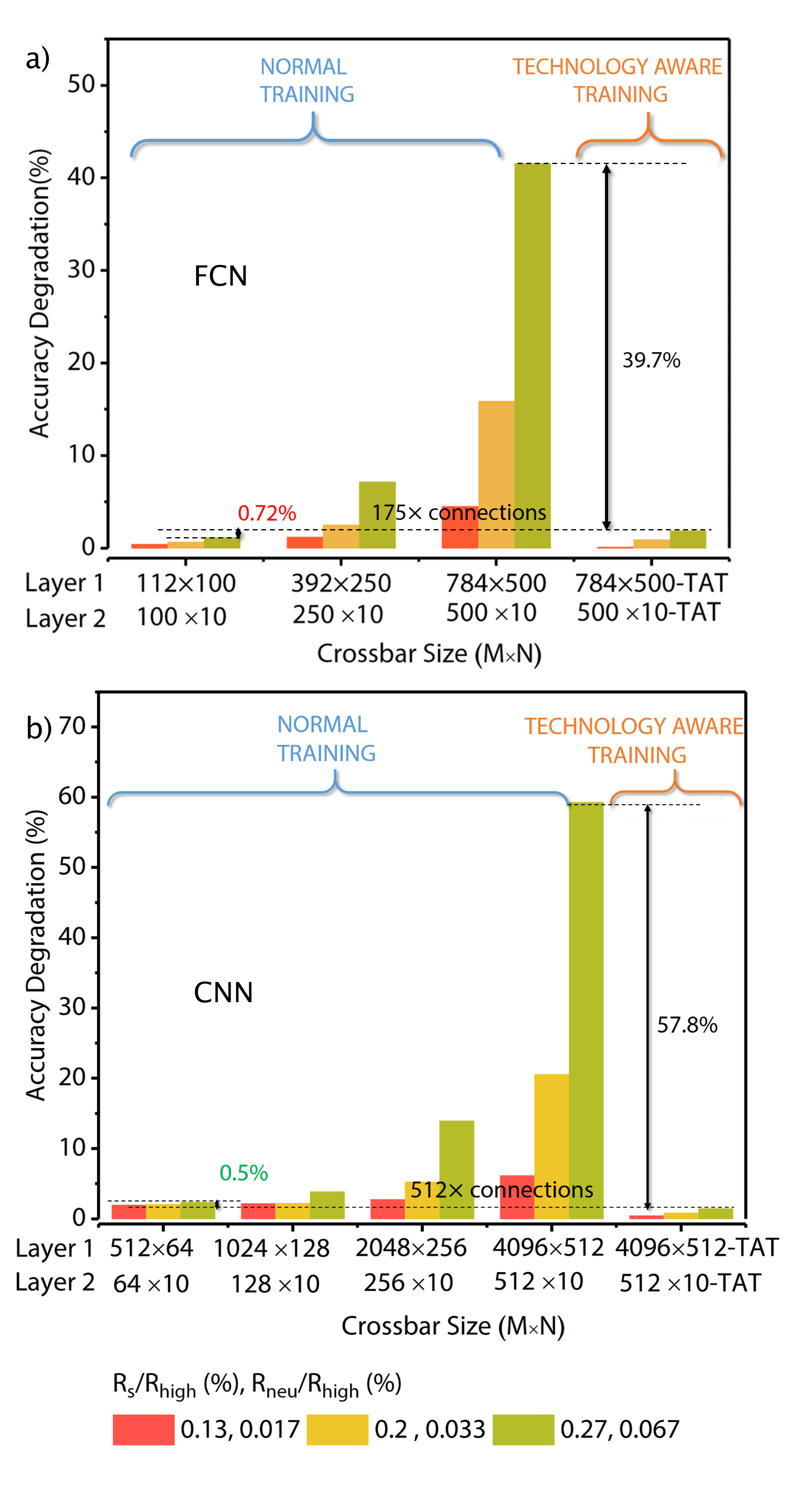}
		\vspace{-8mm}
		\caption{Accuracy degradation v/s crossbar size for various $R_s/R_{high}$ and $R_{neu}/R_{high}$ combinations comparing the technology aware training scheme with normal training for (a) FCN and (b) CNN. Larger crossbars show higher accuracy degradation.}
		
		\label{fig:crossbarvary}
        \vspace{-4mm}
\end{figure}

Non-idealities in crossbars usually establish restrictions on the allowable crossbar sizes due to the dependence of their performance on fan-in and fan-out. For example, the impact of $R_s$ on the crossbar depends on the parallel combination of column resistances and a higher number of columns (and hence, higher fan-out) result in severe performance degradation. Also, the impact of $R_{neu}$ intensifies with increasing number of rows in the crossbar as it leads to more fan-in. As observed in Fig. \ref{fig:crossbarvary}(a), the combined effect of these resistances and variations can result in significant accuracy degradation (41.58\%) when the network is implemented on crossbars of sizes $784\times 500$ and $500 \times 10$ for the respective layers in the FCN. Under the same non-ideal conditions, accuracy degradation drops to 1.2\% when smaller crossbars of sizes $112\times 100$ and $100 \times 10$ are used to represent the functionality of the network. In contrast, considering the same $R_s$ and $R_{neu}$, our proposed training algorithm achieves an accuracy degradation within $\sim 1.89\%$ for sizes $784\times 500, 500\times 10$ and $\sim$ 0.3\% for sizes $112\times 100, 100\times 10$. Thus, the proposed algorithm ensures that a network implemented on larger crossbars can parallel the performance of ideally trained networks implemented on smaller crossbars with minimal degradation. \par
The convolutional layers in CNNs are implemented on smaller crossbars. For the fully connected layers in the CNN architecture, we have considered significantly larger crossbars of sizes $4096\times512$ and $512\times10$. 
Due to large sizes of the last 2 layers of the considered architecture, we show in Fig. \ref{fig:crossbarvary}(b), that the network, when trained under ideal conditions, can suffer as large as $59.3\%$ degradation in accuracy for the worst case resistance constraints considered. On the other hand, using smaller crossbars of sizes  $512\times64, 64\times10$ reduces the accuracy degradation to $2.4\%$ for the same conditions. In comparison, a network trained with the proposed technology aware training algorithm restores the accuracy to within $\sim 1.5\%$ of the ideal accuracy even for the highest crossbar sizes ($4096\times512,512\times10$). Thus, the proposed algorithm ensures that a CNN with fully connected layers implemented on crossbars of size in the order of $4096\times 512$ can achieve better performance than for crossbars of size $512\times 64$ with standard training algorithms. Such a provision of using large crossbars for implementing neuromorphic systems could potentially reduce overheads of repeating inputs, time multiplexing outputs, thus ensuring faster operations.    \par

\section{Conclusion}
Hardware implementations of neuromorphic systems in crossbar architecture could suffer from various non-idealities resulting in severe performance degradation when employed in machine learning applications such as recognition tasks, natural language processing, etc. In this work, we analyzed, by  means of mathematical modeling, the impact of non-idealities such as source resistance, neuron resistance and chip-to-chip device variations on performance of a 3-layered FCN on MNIST and a state-of-the-art CNN architecture on CIFAR-10. Severe degradation in recognition accuracy, up to 59.84\%, was observed in FCNs. Although convolution layers in CNN can be implemented on smaller crossbars, the large fully connected layers at the end made them prone to performance degradation (up to 62.4\% for our example). As a solution, we proposed a technology aware training algorithm which incorporates the mathematical models of the non-idealities in the training algorithm. Considering relevant ranges of non-idealities, our proposed methodology recovered the performance of the network implemented on non-ideal crossbars to within ~2.34\% of the ideal accuracy for FCNs and ∼ 1.5\% for CNNs. We further show that the proposed technology aware training algorithm enables the use of larger crossbars of sizes in the order of $4096\times 512$ for CNNs and $784\times 500$ for FCNs without significant performance degradation. Thus, we believe that the proposed work potentially paves the way for implementation of neuromorphic systems on large crossbars which otherwise is rendered unfeasible using standard training algorithms.


%

\section*{Acknowledgment}
The research was funded in part by the National Science Foundation, Center for Spintronics funded by DARPA and SRC, Intel Corporation, ONR MURI program, and Vannevar Bush Faculty Fellowship.

\bibliographystyle{IEEEtran}
\bibliography{tat1.bib}

\end{document}